# Integrative Pan-Cancer Analysis of RNMT: a Potential Prognostic and Immunological Biomarker

Shuqiang Huang[1], Cuiyu Tan[1], Jinzhen Zheng[1], Zhugu Huang[2], Zhihong Li[1], Ziyin Lv[3], Wanru Chen[4]

[1]The Sixth Affiliated Hospital of Guangzhou Medical University, Qingyuan, Guangdong 511500, China.

[2]School of Pediatrics, Guangzhou Medical University, Guangzhou, Guangdong 510000, China

[3]The First Clinical College of Guangzhou Medical University, Guangzhou, Guangdong 511436, China.

[4]The Third Clinical College of Guangzhou Medical University, Guangzhou, Guangdong 511436, China.

ABSTRACT

**Background:** RNA guanine-7 methyltransferase (RNMT) is one of the main regulators of N7-methylguanosine (m7G), and the deregulation of RNMT correlated with tumor development and immune metabolism. However, the specific function of RNMT in pan-cancer remains unclear.

**Methods:** RNMT expression in different cancers was analyzed using multiple databases, including Cancer Cell Line Encyclopedia (CCLE), Genotype-Tissue Expression Project (GTEx), and The Cancer Genome Atlas (TCGA). Cox regression analysis and Kaplan-Meier analysis were used to estimate the correlation of RNMT expression to prognosis. The data was also used to research the relationship between RNMT expression and common immunoregulators, tumor mutation burden (TMB), microsatellite instability (MSI), mismatch repair (MMR), and DNA methyltransferase (DNMT). Additionally, the cBioPortal website was used to evaluate the characteristics of RNMT alteration. The TISDB database was used to obtain the expression of different subtypes. The Tumor Immune Estimation Resource (TIMER) database was used to analyze the association between RNMT and tumor immune infiltration. Gene set enrichment analysis (GSEA) was used to identify the relevant pathways.

**Results:** RNMT was ubiquitously highly expressed across cancers and survival analysis revealed that its expression was highly associated with the clinical prognosis of various cancer types. Remarkably, RNMT participates in immune regulation and plays a crucial part in the tumor microenvironment (TME). A positive association was found between RNMT expression and six immune cell types expression (B cell, CD4 + T cell, CD8 + T cell, dendritic, macrophage, and neutrophil) in colon adenocarcinoma (COAD), kidney renal clear cell carcinoma (KIRC), and liver hepatocellular carcinoma (LIHC). Moreover, RNMT expression was highly associated with immunoregulators in most cancer types, and correlated to TMB, MSI, MMR, and DNMT. Finally, GSEA indicated that RNMT may correlate with tumor immunity.

**Conclusion:** RNMT was ubiquitously related to tumor prognosis and tumor immune infiltrates, and played an essential role in the occurrence and progression of various tumors. RNMT may be a promising prognostic biomarker in various tumors and provide new ideas for future tumor immune studies and treatment strategies.

## INTRODUCTION

It is an intimidating and everlasting struggle for the human race to fight cancer. As the framework of cancer diversity has been constantly polished (1), searching for powerful tumor markers offers a promising avenue for individual-based tumor diagnosis and treatment. Over the last few decades, modern immunotherapy, as a novel and promising direction for cancer treatment, has made considerable progress and revolutionized the therapy of tumors in many ways. At present, there are a variety of immunotherapy strategies including the use of monoclonal antibodies, adjuvants, checkpoint inhibitors and so on (2). In the case of immune inhibitors, Immune checkpoints refer to numerous inhibitory pathways hardwired into the immune system that are essential for the maintenance of self-tolerance and the prevention of autoimmunity (3). In practical clinical applications, the blockade of immune checkpoints is one of the most promising approaches to activating therapeutic antitumor immunity. For example, combined PD-1 and CTLA-4 pathway inhibition approaches demonstrated increased survival markedly in multiple malignancies, including melanoma, renal cell carcinoma, esophageal squamous-cell carcinoma, and non–small cell lung cancer (4-7). Driven by such promising results, pre-clinical research on checkpoint inhibitors has become a rapidly evolving field. Nonetheless, there are still many challenges in cancer immunotherapy, which range from a lack of confidence in translating pre-clinical findings to confirming the best combination for any given patient (8). Therefore, more clinical and translational research is undoubtedly needed to explore potential and optimal targets for cancer immunotherapy.

N7-methylguanosine (m7G), occurring in the N7-position of guanosine, is one of the electropositive modifications of messenger RNA (mRNA) 5′ cap structure, internal mRNA, ribosomal RNA (rRNA), transfer RNA (tRNA), microRNA (miRNA), and long non-coding RNAs (lncRNA) (9-11). The m7G modification plays a significant role in mRNA export, metabolism, stability, splicing, and translation (12). Recent evidence suggests that the change of m7G has a great effect on the development of cancer, which means that m7G molecules may be potential therapeutic targets and prognostic biomarkers (13). RNA guanine-7 methyltransferase (RNMT), localized on chromosome 18p11.21, contains an N-terminal regulatory domain that regulates RNMT activity and facilitates recruitment to transcription initiation sites (14, 15). RNMT/RNMT-activating miniprotein (RAM)-mediated m7G mRNA methylome modification at the 5' caps is important for ordinary RNA translation, the transformation of mammary epithelial and fibroblast cell, and T cell activation (16, 17). Previous research has established that aberrant expression of RNMT is also closely associated with a variety of cancers. For example, the promoters of RNMT are hypomethylated in hepatocellular carcinoma tumorigenesis, which promotes the development of HCC (18). Inhibition of RNMT induces the apoptosis of HeLa cells, while decreasing the growth of breast cancer cells in a PIK3CA mutant background (19, 20). Moreover, the expression of RNMT is decreased in glioma stem-like cells (GSLCs) and inhibits tumor growth via the B7-H6/c-Myc axis (21). Although the function of RNMT in several tumors has been explored, its key functions in tumor immunity are still obscure.

With the rise of bioinformatics, multiple datasets in existence can be used to forecast the effect of numerous genes. In this research, we completed Pan-cancer data collection and processing by looking for the databases such as The Cancer Genome Atlas (TCGA) database, the Genotype-Tissue Expression (GTEx) database, and The Tumor Immune Estimation Resource（TIMER） database. Afterwards, we make analyses of survival and immune prognosis, TMB, MSI, DNA MMR Genes, DNMT and Gene Set Enrichment Analysis using a variety of statistical methods, including Cox regression and Kaplan–Meier analysis, log-rank statistic, spearman correlation analysis and so on to validate the value of RNMT in pan-cancer indicative function.

**MATERIALS AND METHODS**

**Sample Source Obtained and the Expression of RNMT in Pan-Cancer**
In this study, a series of standardized pan-cancer data were downloaded from The Cancer Genome Atlas(TCGA) database （https://portal.gdc.cancer.gov）, the Genotype-Tissue Expression (GTEx) database(https://commonfund.nih.gov/GTEx/)and Cancer Cell Line Encyclopedia (CCLE) database(https://portals.broadinstitute.org/ccle). Furthermore, we extract the expression information of RNMT in different samples so as to identify the role of RNMT in Pan-cancer. And the abbreviation and the full name of all the cancers are shown in **Table 1**.

**TABLE 1 Abbreviations of the tumors from TCGA database**

| Abbreviations | Tumor name |
|---|---|
| ACC | Adrenocortical carcinoma |
| BLCA | Bladder Urothelial Carcinoma |
| BRCA | Breast invasive carcinoma |
| CESC | Cervical squamous cell carcinoma and endocervical adenocarcinoma |
| CHOL | Cholangiocarcinoma |
| COAD | Colon adenocarcinoma |
| COADREAD | Colon and rectal cancer |
| DLBC | Lymphoid Neoplasm Diffuse Large B-cell Lymphoma |
| ESCA | Esophageal carcinoma |
| FPPP | FFPE Pilot Phase II FFPE |
| GBM | Glioblastoma multiforme |
| GBMLGG | Glioma |
| HNSC | Head and Neck squamous cell carcinoma |
| KICH | Kidney Chromophobe |
| KIPAN | Pan-kidney cohort (KICH+KIRC+KIRP) |
| KIRC | Kidney renal clear cell carcinoma |
| KIRP | Kidney renal papillary cell carcinoma |
| LAML | Acute Myeloid Leukemia |
| LGG | Brain Lower Grade Glioma |
| LIHC | Liver hepatocellular carcinoma |
| LUAD | Lung adenocarcinoma |
| LUSC | Lung squamous cell carcinoma |
| MESO | Mesothelioma |
| OV | Ovarian serous cystadenocarcinoma |
| PAAD | Pancreatic adenocarcinoma |
| PCPG | Pheochromocytoma and Paraganglioma |
| PRAD | Prostate adenocarcinoma |
| READ | Rectum adenocarcinoma |
| SARC | Sarcoma |
| SKCM | Skin Cutaneous Melanoma |
| STAD | Stomach adenocarcinoma |
| STES | Stomach and Esophageal carcinoma |
| TGCT | Testicular Germ Cell Tumors |
| THCA | Thyroid carcinoma |
| THYM | Thymoma |
| UCEC | Uterine Corpus Endometrial Carcinoma |
| UCS | Uterine Carcinosarcoma |
| UVM | Uveal Melanoma |

**Cox Regression and Kaplan–Meier Prognosis Analysis**

The RNMT expression data of normal samples from GTEx and cancer samples from TCGA were extracted and formed into an expression matrix. Univariate cox model was used to reveal the relationship between RNMT expression and patient survival. On the basis of the best separation of expression of RNMT, the Kaplan-Meier (K-M) method by log-rank test was used to compare the prognosis of patients (overall survival: OS; disease-specific survival: DSS; disease-free interval: DFI; and progression-free interval: PFI) from the high and low group. The images were plotted as forest plots (Cox regression analysis) and K-M curves (K-M prognosis analysis).

**Genetic Alteration Analysis of RNMT in Human Pan-Cancer**

The cBioPortal website (http://www.cbioportal.org) provides a platform for analyzing, visualizing, and exploring multidimensional cancer genomic data (22). The database was used to analyze the characteristics of genetic alteration, including missense mutation, deep deletion, copy number amplification, and RNMT mRNA upregulation.

**Immune and Molecular Subtype Analysis**

In order to have a better understanding of the relationship between immune and molecular subtypes and RNMT expression among human cancers, the TISDB database (http://cis.hku.hk/TISIDB/index.php) was used to obtain the analysis result and determine the expression of different subtypes. It was considered statistically significant when the $p < 0.05$.

**Immune Infiltration and Tumor Microenvironment Assess**

The Tumor Immune Estimation Resource (TIMER) database (http://cistrome.dfci.harvard.edu/TIMER/) is an influential website, which can evaluate the condition of immune infiltration and tumor microenvironment, so we used it to score the RNMT expression in diverse immune cell infiltrations and tumor microenvironment types by the means of the standardized expression matrix.

**Correlation Analysis of RNMT Expression Level and Immunoregulator and Immune Checkpoints**

The TCGA dataset was downloaded to extract the expression data of RNMT gene and immunoregulator (chemokines, receptors, TILs, immunostimulators, and immunoinhibitors), immune checkpoints (24 inhibitors and 36 stimulators). Spearman's correlation test was used to analyze the association between RNMT expression and immunoregulator, immune checkpoints.

**Analysis of Tumor Mutation Burden(TMB), Microsatellite Instability(MSI), Mismatch Repair(MMR) and DNA Methyltransferase(DNMT)**

TMB, MSI, DNA MMR and DNA methylation were considered prognosis-related and immune-related factors in cancer immunotherapy (23). The correlations of RNMT expression with TMB and MSI in human pan-cancer were based on the TCGA pan-cancer atlas. We also analyzed five MMR genes (MLH1, MSH2, MSH6, EPCAM, and PMS2) and four methyltransferases (DNMT1, DNMT2, DNMT3A, and DNMT3B) in human pan-cancer from the TCGA database. In this study, spearman correlation analysis was applied to evaluate the role of RNMT expression in TMB, MSI, MMR, and DNMT.

**Gene Set Enrichment Analysis (GSEA) of RNMT in Human Pan-Cancer**

GSEA is a common method used to interpret gene expression data and compare different groups with differing biological states to determine statistically significant differences (24). The signaling pathway of RNMT was analyzed with the R package clusterProfiler. The Kyoto Encyclopedia of Genes and Genomes (KEGG; https://www.kegg.jp.) was applied. Implementation criteria were normalized enrichment score | NES | > 1, $p < 0.05$, false discovery rate (FDR) ≤ 0.25.

**Statistical Analysis**

Statistical analysis methods were described in the above parts. $p < 0.05$ was recorded as statistically significant for all analyses unless otherwise specified.

# RESULT

Through Pan-Cancer analysis, similarities and differences between genomic and cellular changes of RNMT in different tumor types can be found, and it also plays an important role in excavating potential prognostic markers and therapeutic targets.

## The Expression Level of RNMT in Tissues

Data from GTEx database and CCLE database showed that RNMT was normally expressed in 31 normal tissues, with higher expression levels present in the bone marrow, pituitary, ovary and thyroid, and lower expression levels in blood, pancreas and liver **(Figure 1A)**. Among various cancer cell lines, RNMT was significantly elevated expressed in salivary gland, skin and stomach than in the other tissues, while more lowly expressed in kidney and breast **(Figure 1B)**.

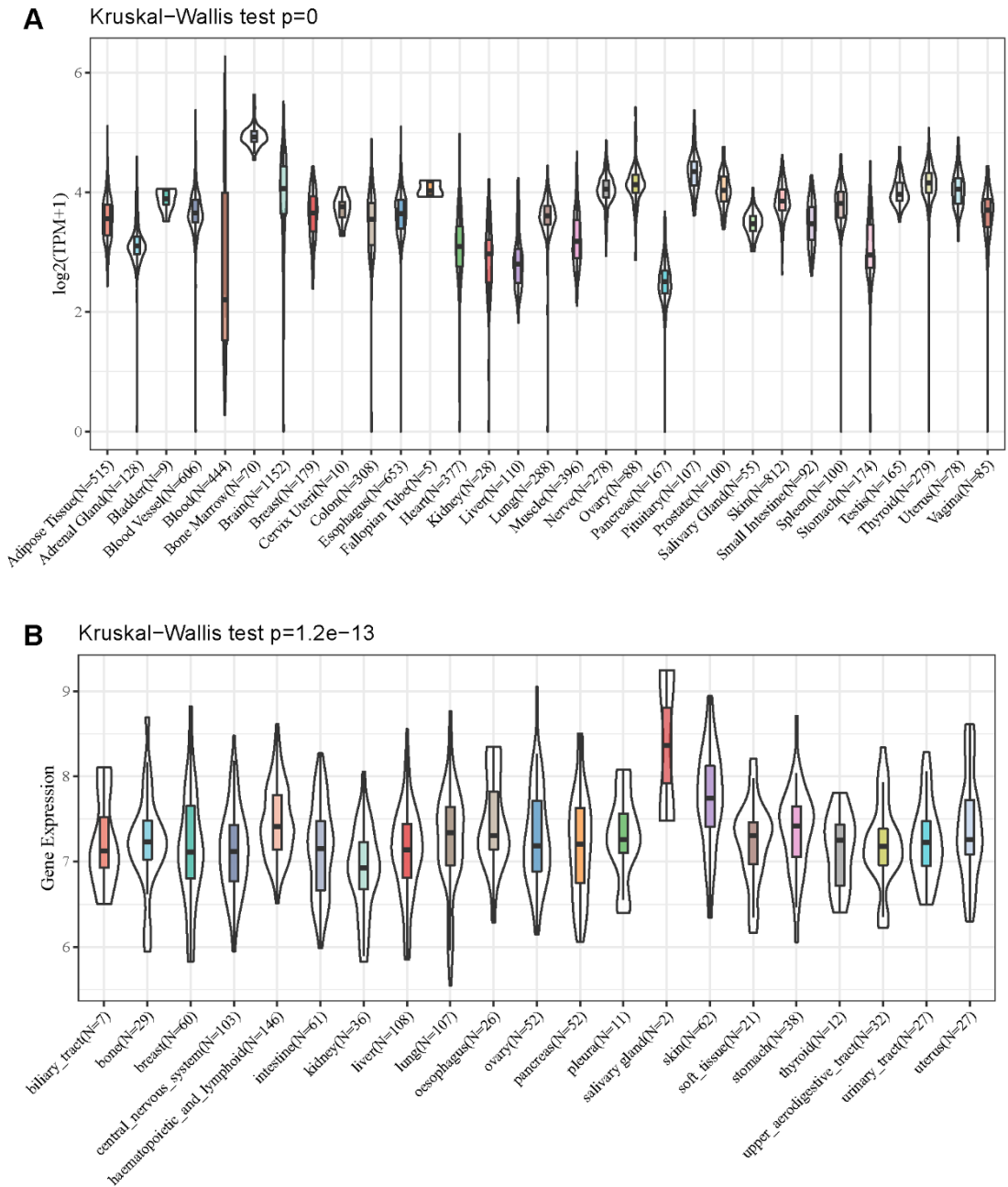

**FIGURE 1** The GTEx and CCLE databases showed that RNMT is expressed in a wide range of tissues at varying levels. **(A)** RNMT expression in 31 normal tissues from the GTEx database. **(B)** RNMT expression in 21 cancer cell lines from the CCLE database.

As focused on analyzing the expression pattern of RNMT in normal tissues and corresponding tumors, we found that the expression of RNMT in 15 types of tumors was having increased, including LUAD, COAD, COADREAD, ESCA, STES, KIRP, KIPAN, STAD, HNSC, KIRC, LUSC, LIHC, PCPG, BLCA and CHOL, while it was dramatically decreased in GBM, GBMLGG, and LGG (based on TCGA database). Obtained results from GTEx database and TCGA database, there are 20 types of tumors upregulating RNMT expression significantly compared to the normal tissues samples, including GBMLGG, LGG, BRCA, ESCA, STES, KIRP, KIPAN, COADREAD, STAD, HNSC, KIRC, LUSC, LIHC, SKCM, PAAD, UCS, LAML, PCPG, ACC and CHOL. In contrast, it was reported that RNMT has lower expressed in GBM, COAD, PRAD, OV and TGCT relative to the normal tissues.

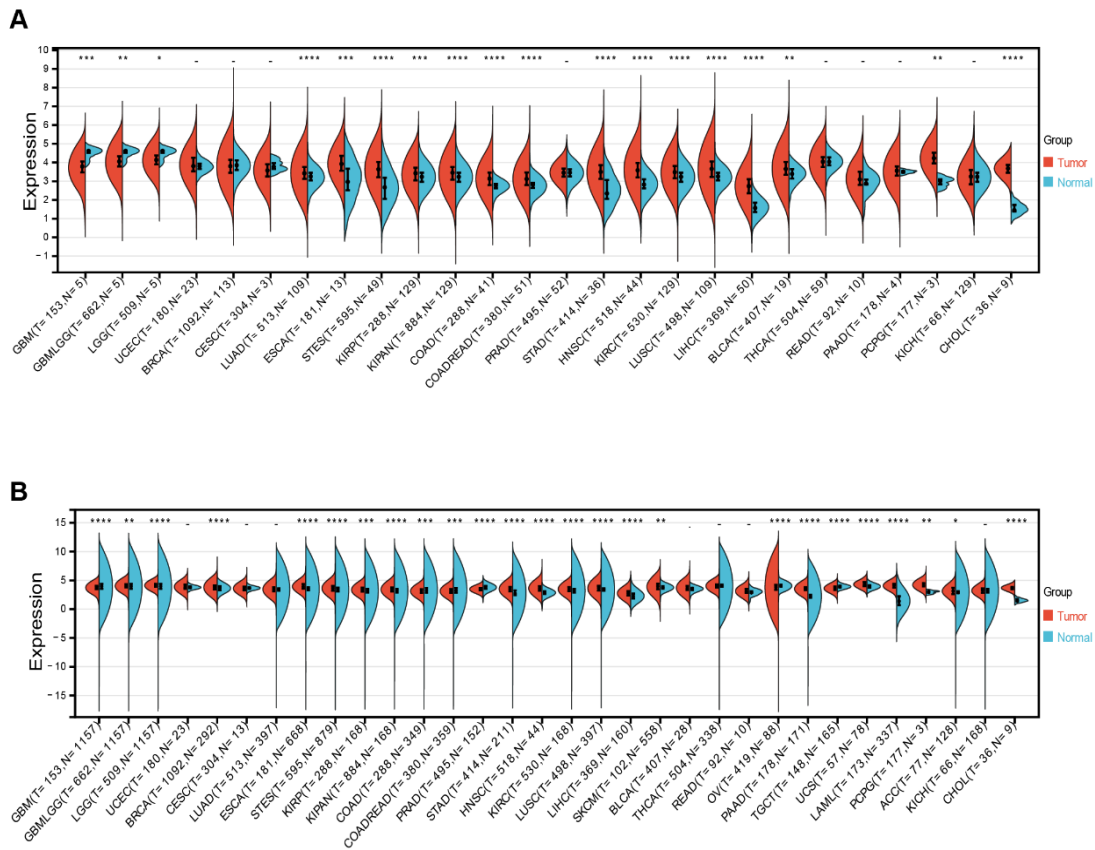

**FIGURE 2** The TCGA and GTEx database showed that RNMT is expressed in a wide range of tissues at varying levels. **(A)** Differential expression pattern of RNMT in normal tissues and corresponding tumor from TCGA database. **(B)** Differential expression pattern of RNMT in normal tissues and corresponding tumor from TCGA database and GTEx database. (*$p < 0.05$, **$p < 0.01$, ***$p < 0.001$ and ****$p < 0.0001$).

**Prognostic Value of RNMT in Human Pan-Cancer**

To explore the relationship between RNMT and the clinical prognosis of pan-cancer patients, datasets from TCGA target GTEx were used for survival association analysis. Cox regression model showed that RNMT expression impacted overall survival (OS) in 7 cancer types, including LIHC ($p = 0.01$, HR = 1.41), ACC ($p = 0.02$, HR = 1.81), KICH ($p = 0.0033$, HR = 7.12), GBMLGG ($p < 0.0001$, HR = 0.55), THYM ($p = 0.03$, HR = 0.48), SKCM ($p = 0.05$, HR = 0.84), and OV ($p = 0.00085$, HR = 0.83) **(Figure 3A)**. The Kaplan–Meier (K-M) curves revealed high expression of RNMT correlates with poor OS in LIHC ($p = 0.00055$), ACC ($p = 0.0045$), KICH ($p = 0.00052$), while high expression of RNMT correlates with good OS in GBMLGG ($p < 0.0001$), THYM ($p = 0.01$), and OV ($p = 0.0082$) **(Figure 3B-G)**.

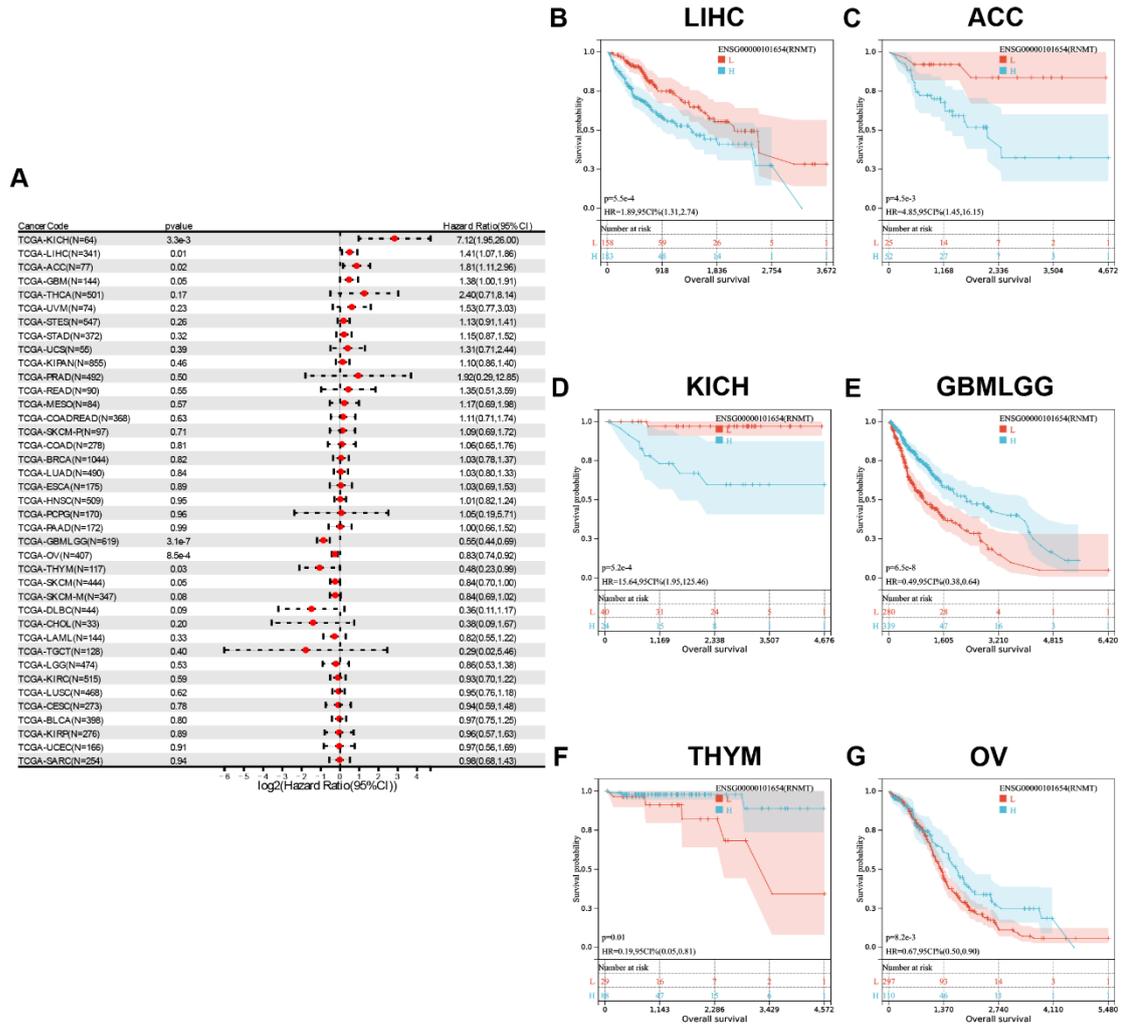

**FIGURE 3** The correlation between RNMT expression and overall survival (OS). **(A)** Cox regression analysis of RNMT with OS in different cancer types. **(B-G)** Kaplan-Meier survival analysis of RNMT expression and OS in LIHC, ACC, KICH, GBMLGG, THYM, and OV.

However, it is possible that OS could be adversely affected by deaths other than cancer during follow-up. Thus, we reanalyzed the data in order to study a correlation between DSS and RNMT expression among different tumor types. The results of Cox analysis showed that RNMT expression impacted DSS in 6 cancer types, including GBM ($p = 0.02$, HR = 1.54), LUSC ($p = 0.03$, HR = 1.47)、ACC ($p = 0.01$, HR = 1.94), KICH ($p = 0.01$, HR = 6.69), GBMLGG ($p < 0.0001$, HR = 0.53), OV ($p = 0.00029$, HR = 0.81) **(Figure 4A)**. The K-M curves displayed that high expression of RNMT correlates with poor DSS in ACC ($p = 0.003$), KICH ($p = 0.0032$), while high expression of RNMT correlates with good DSS in GBMLGG ($p < 0.0001$) and OV ($p = 0.0024$) **(Figure 4B-E)**.

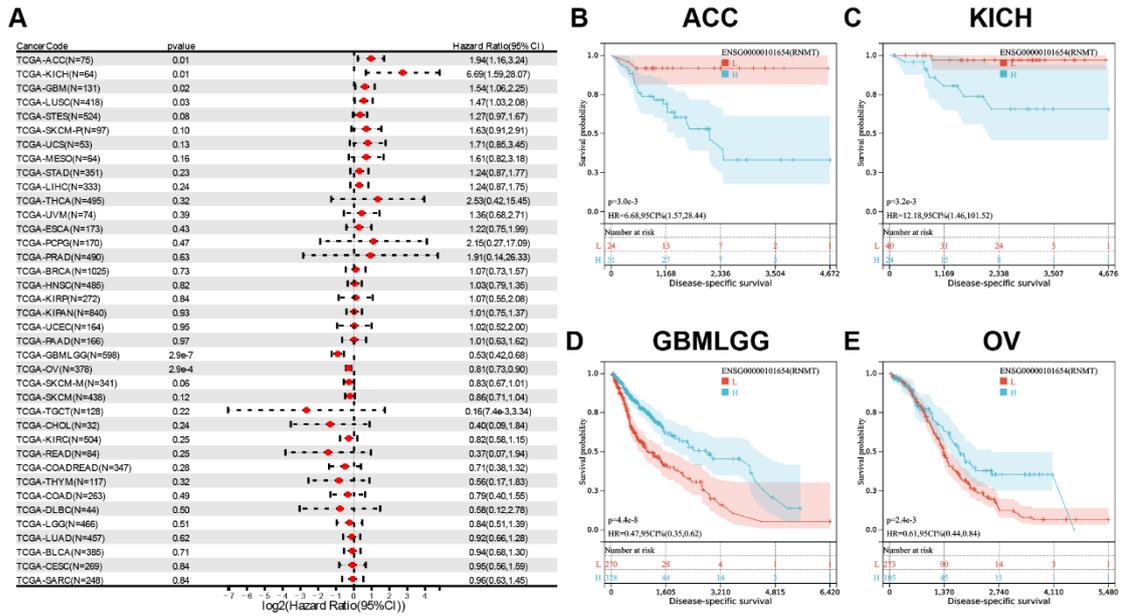

**FIGURE 4** The correlation between RNMT expression and disease-specific survival (DSS). **(A)** Cox regression analysis of RNMT with DSS in different cancer types. **(B-E)** Kaplan-Meier survival analysis of RNMT expression and DSS in ACC, KICH, GBMLGG, and OV.

Moreover, the correlation of RNMT expression with patients' DFI was also analyzed. Cox regression model depicted the relationship in PRAD ($p = 0.03$, HR = 3.37), LUSC ($p = 0.04$, HR = 1.59), ACC ($p = 0.0019$, HR = 3.25), and OV ($p = 0.02$, HR = 0.74) **(Figure 5A)**. The K-M curves showed that high expression of RNMT correlates with poor DFI in PRAD ($p = 0.01$), LUSC ($p = 0.0062$), ACC ($p = 0.0031$), while high expression of RNMT correlates with good DFI in OV ($p = 0.0071$) **(Figure 5B-E)**.

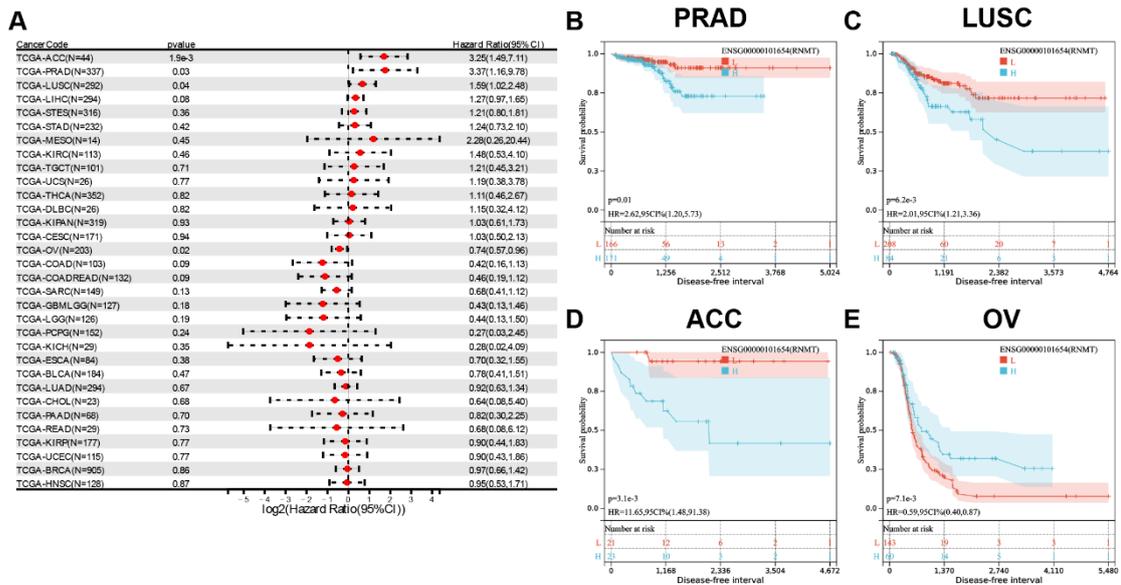

**FIGURE 5** The correlation between RNMT expression and disease-free interval (DFI). **(A)** Cox regression analysis of RNMT with DFI in different cancer types. **(B-E)**

Kaplan-Meier survival analysis of RNMT expression and DFI in PRAD, LUSC, ACC, and OV.

Finally, regarding associations between RNMT expression and PFI, Cox regression analysis showed that RNMT expression had a correlation with 7 cancer types, including STES ($p = 0.04$, HR=1.25), LUSC ($p = 0.0087$, HR=1.44), LIHC ($p = 0.02$, HR = 1.32), ACC ($p < 0.0001$, HR = 2.28), GBMLGG ($p < 0.0001$, HR = 0.51), SKCM-M ($p = 0.02$, HR = 0.83), and OV ($p = 0.02$, HR = 0.87) **(Figure 6A)**. In addition, the K-M PFI curves showed that high expression of RNMT correlates with poor prognosis in STES ($p = 0.03$), LUSC ($p = 0.02$), LIHC ($p = 0.00021$), ACC ($p < 0.0001$), while high expression of RNMT correlates with good prognosis in GBMLGG ($p < 0.0001$), SKCM-M ($p = 0.03$), and OV ($p = 0.04$) **(Figure 6B-H)**.

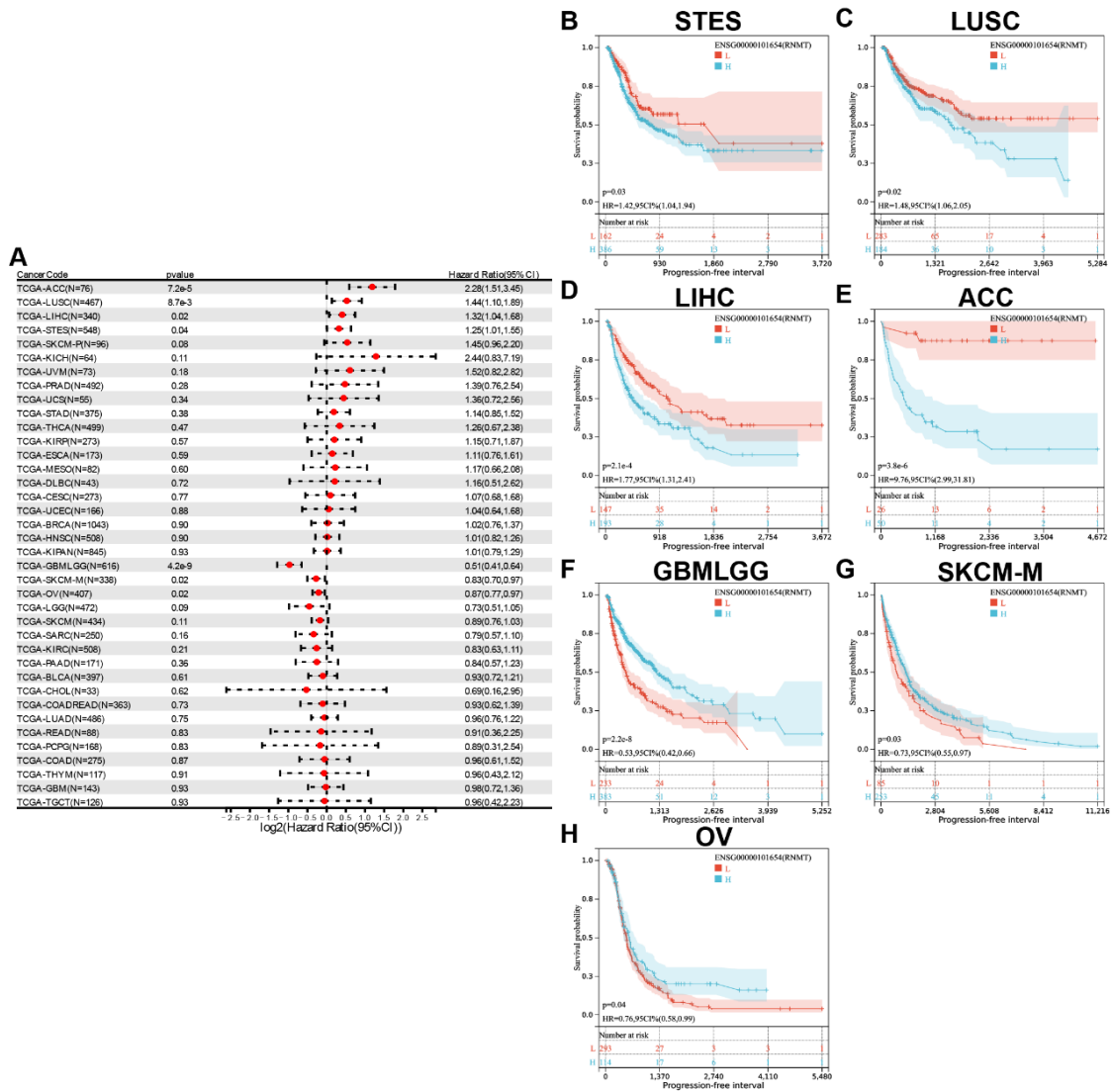

**FIGURE 6** The correlation between RNMT expression and progression-free interval (PFI). **(A)** Cox regression analysis of RNMT with PFI in different cancer types. **(B-H)** Kaplan-Meier survival analysis of RNMT expression and PFI in STES, LUSC, LIHC, ACC, GBMLGG, SKCM-M, and OV.

## RNMT Gene Mutation

A genomic mutation influences tumor development. Using the cBioPortal database, we analyzed the characteristics of RNMT genetic alterations **(Figure 7).** As a result, RNMT mutation was seen in the following tumor tissues as one of the most crucial factors: UCEC, STAD, SKCM, DLBC, LUSC, LUAD, CESC, COAD, PRAD, KIRP, GBM, KIRC, PCPG, and THCA. The amplification of RNMT is also one of the most crucial factors due to its alteration in the following tumor tissues: BLCA, PAAD, LAML, UCS, ESCA, HNSC, OV, BRCA, LIHC, SARC, and LGG.

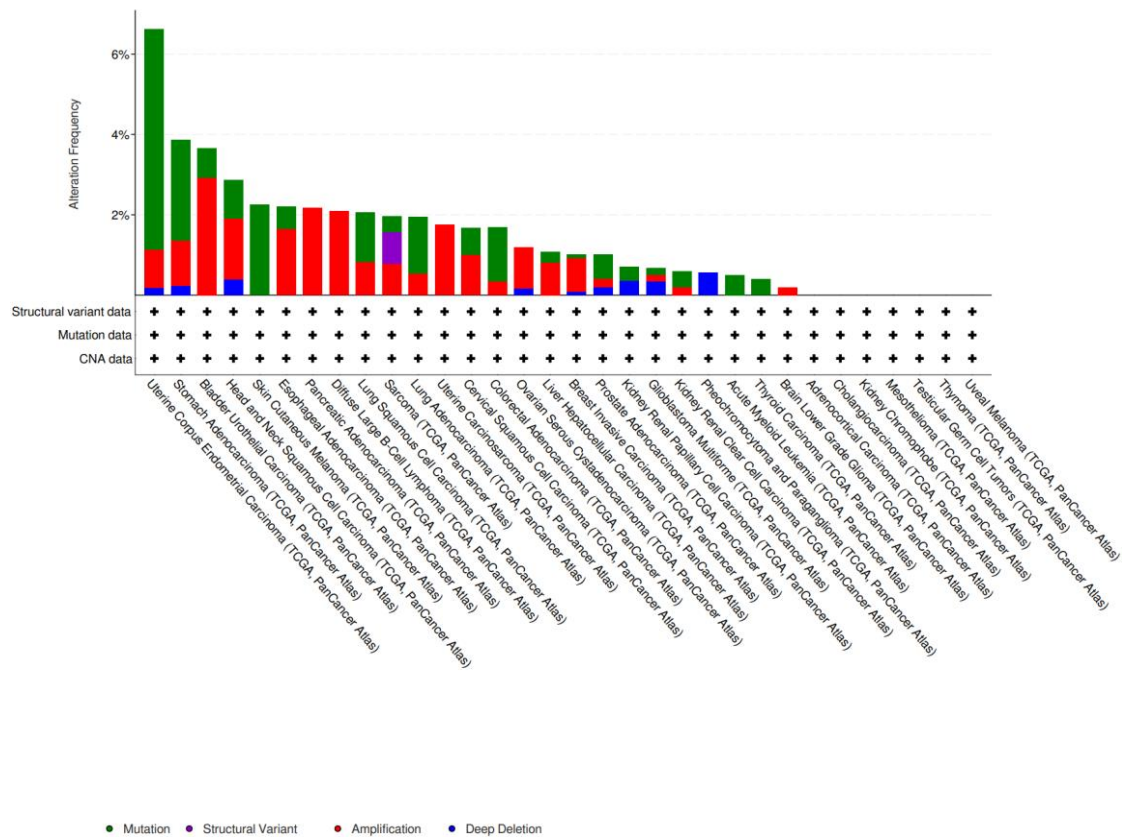

**FIGURE 7** RNMT genomic alterations analyzed by the cBioPortal database.

## Correlations Between RNMT Expression and Immune and Molecular Subtypes

Cancer Immune Landscape consists of six subtypes, including C1 (wound healing), C2 (IFN-1 dominant), C3 (inflammatory), C4 (lymphocyte depleted), C5 (immunologically quiet), and C6 (TGF-β dominant), which was determined by the distinction of macrophage or lymphocyte signatures (25). And it has a promising perspective in prognostic analysis and targeted therapy. The analysis result of TISDB database showed that RNMT was significantly different in ACC, CESC, HNSC, KIRC, LGG, LUSC, OV, PRAD, SKCM, TGCT, UCEC, and LGG (the most remarkably differential with RNMT) was presented six kinds of immune subtypes(C3, C4, C5 and C6) **(Figure 8).**

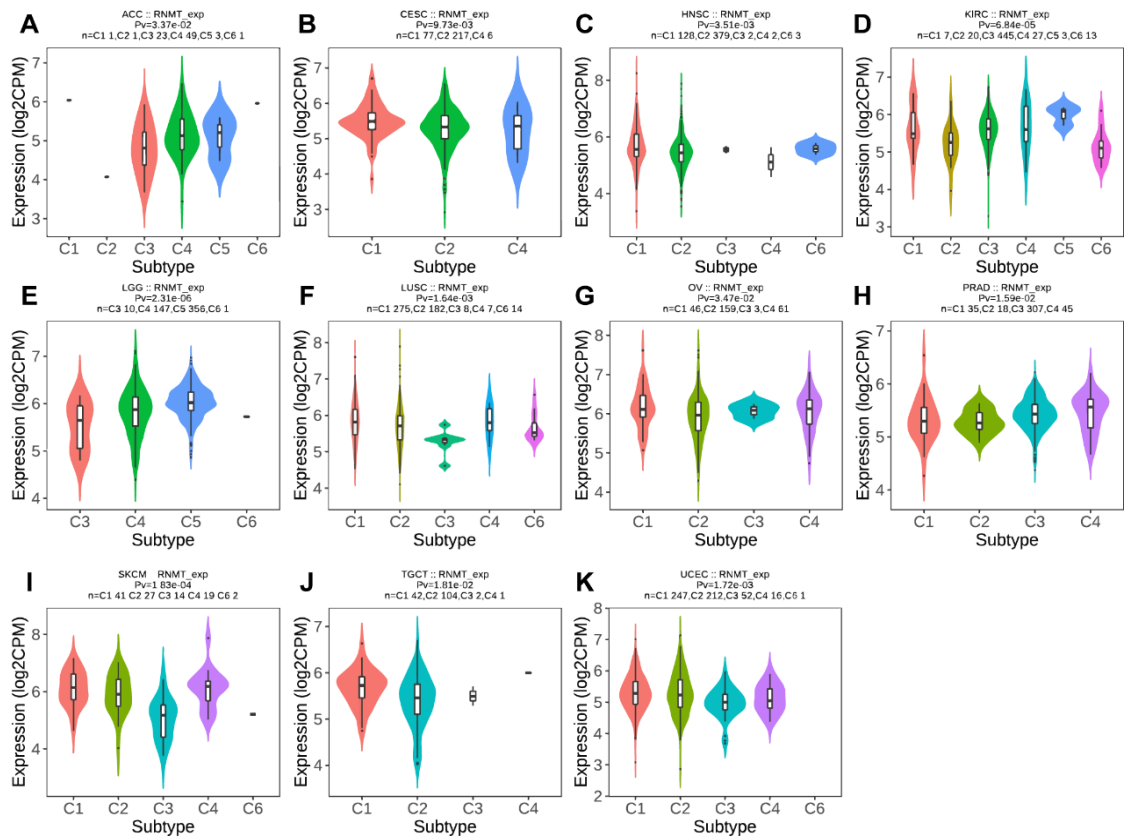

**FIGURE 8** Correlations between various immune subtypes and RNMT expression in pan-cancer cells, note: C1 (wound healing), C2 (IFN-1 dominant), C3 (inflammatory), C4 (lymphocyte depleted), C5 (immunologically quiet), and C6 (TGF-β dominant).

For various molecular subtypes of cancers, RNMT was significantly different in BRCA, COAD, HNSC, LGG, LUSC, OV, PCPG, PRAD, READ, STAD, and UCEC **(Figure 9)**. The results indicate that RNMT expression differs in immune and molecular subtypes of diverse types of cancer.

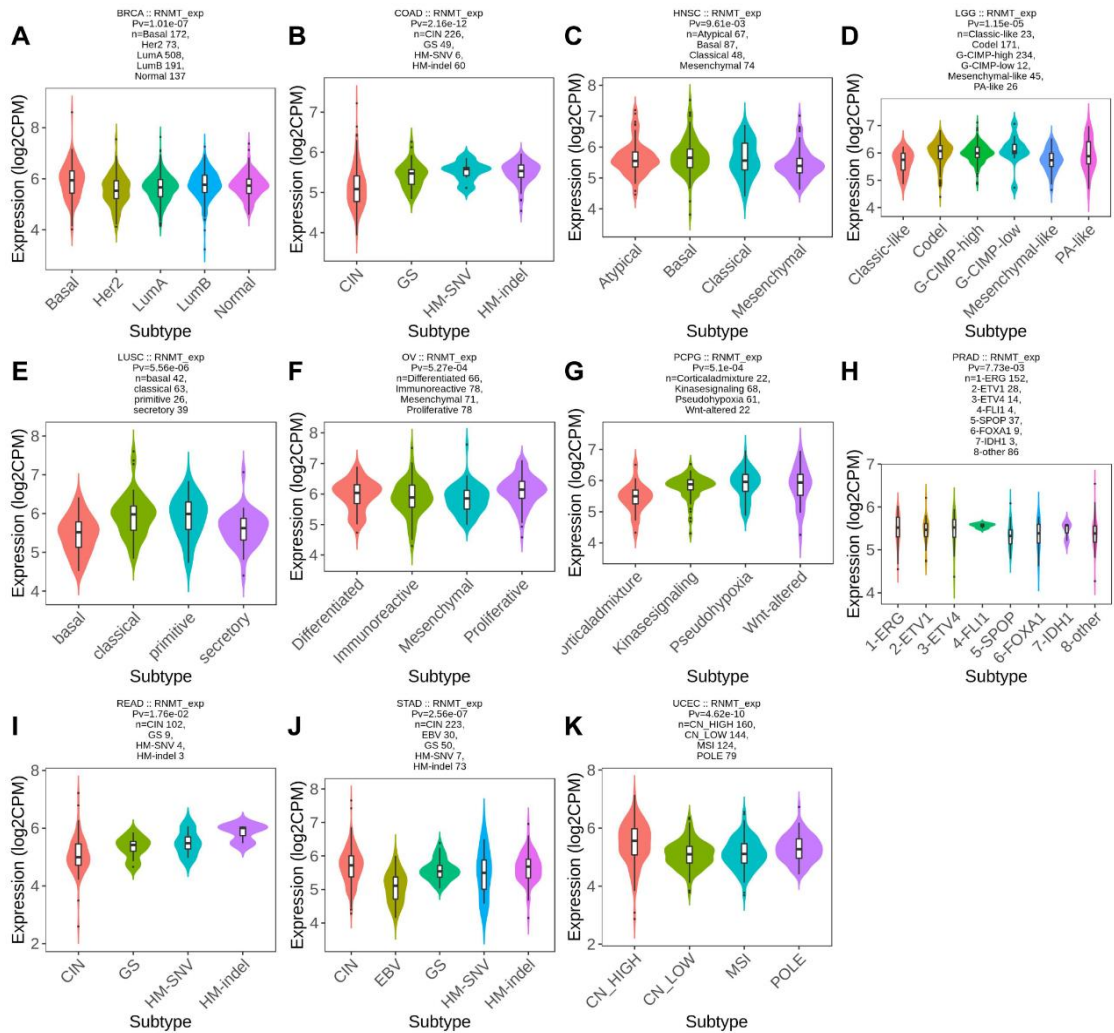

**FIGURE 9** Correlations between various molecular subtypes and RNMT expression in pan-cancer cells.

## Correlations Between RNMT and Tumor Immune Infiltration and Microenvironment in Human Pan-Cancer

Scoring the correlation between RNMT and Immune cell infiltration, as well as tumor microenvironment aids to predict prognosis and treatment sensitivity in cancers (26). Exploiting the TIMER database to contrast the relationship between RNMT and immune cell infiltration with different cancers, RNMT had the strongest correlation with 3 tumors, including COAD, KIRC, and LIHC. And they had one thing in common that upregulated RNMT expression is associated with increased six immune cell types expression (B cells, CD4+ T cell, CD8+ T cell, dendritic, macrophages and neutrophils). In addition, the expression of the immune cells in COAD contributed to a stronger positive correlation with RNMT than in any other tumors. More detailed information of the result was shown in **Figure 10A**.

In tumor microenvironment analysis, we ranked 31 types of tumors according to the P value and only the top three were presented here. COAD (R = 0.27, $p < 0.001$) was found to be a positive correlation with RNMT while SARC (R = -0.333, $p < 0.001$) and LUSC (R = -0.19, $p < 0.001$) were negatively correlated with the expression levels of RNMT in Stromal score. Moreover, three of the most significant correlation about immune score (SARC, UCEC and LUSC) had a negative association with the gene RNMT, since the similar result was obtained in the estimate immune score between the expression of RNMT and SARC, LUSC and UCEC **(Figure 10B)**.

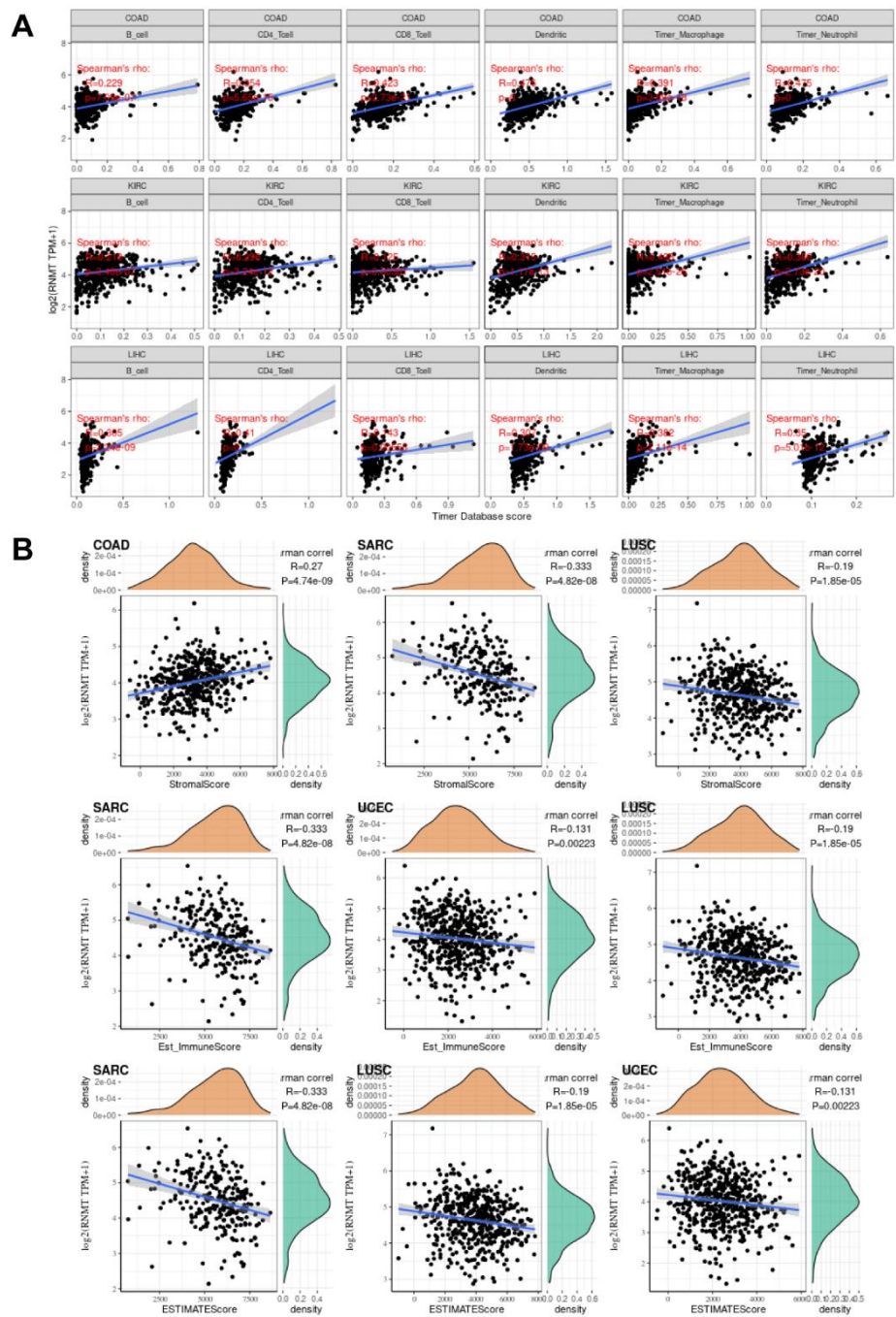

**FIGURE 10 (A)** Relationship of RNMT expression level with immune cell infiltration in COAD, KIRC and LIHC. **(B)** Connection of RNMT expression with stromal score, immune score, and estimate immune score.

**Relationship Between RNMT Expression and Immunoregulators in Human Pan-Cancer**

The relationship between RNMT and immunomodulators was shown in **Figure 11**. RNMT expression was found to be positively correlated with the majority of immunoregulators in COADREAD, COAD, UVM, KIPAN, and PRAD. Furtherly, it was found that RNMT was positively correlated with immunostimulator CD 276 and TNFRSF13C expressions and immunoinhibitor TGFBR1 expression in most cancers. Due to immunotherapies is a considerable treatment for tumor reduction and eradication, the correlation analysis of RNMT expression and immune checkpoints was performed, including 24 immune inhibitors and 36 stimulators. Our results showed that RNMT expression levels correlated with many immune checkpoints in most of the cancer types, with a positive correlation especially in UVM, COAD, COADREA, PRAD, KIPAN, and LIHC, a negative association mainly in GBMLGG, SARC, TGCT, LGG, and THCA. The results also displayed that RNMT was positively correlated with immune checkpoints including CD276, VEGFA, and HMGB1 in multiple cancers **(Figure 12)**. These results suggested that RNMT has the ability to modulate these immune checkpoints and may strengthen immunity in some tumors.

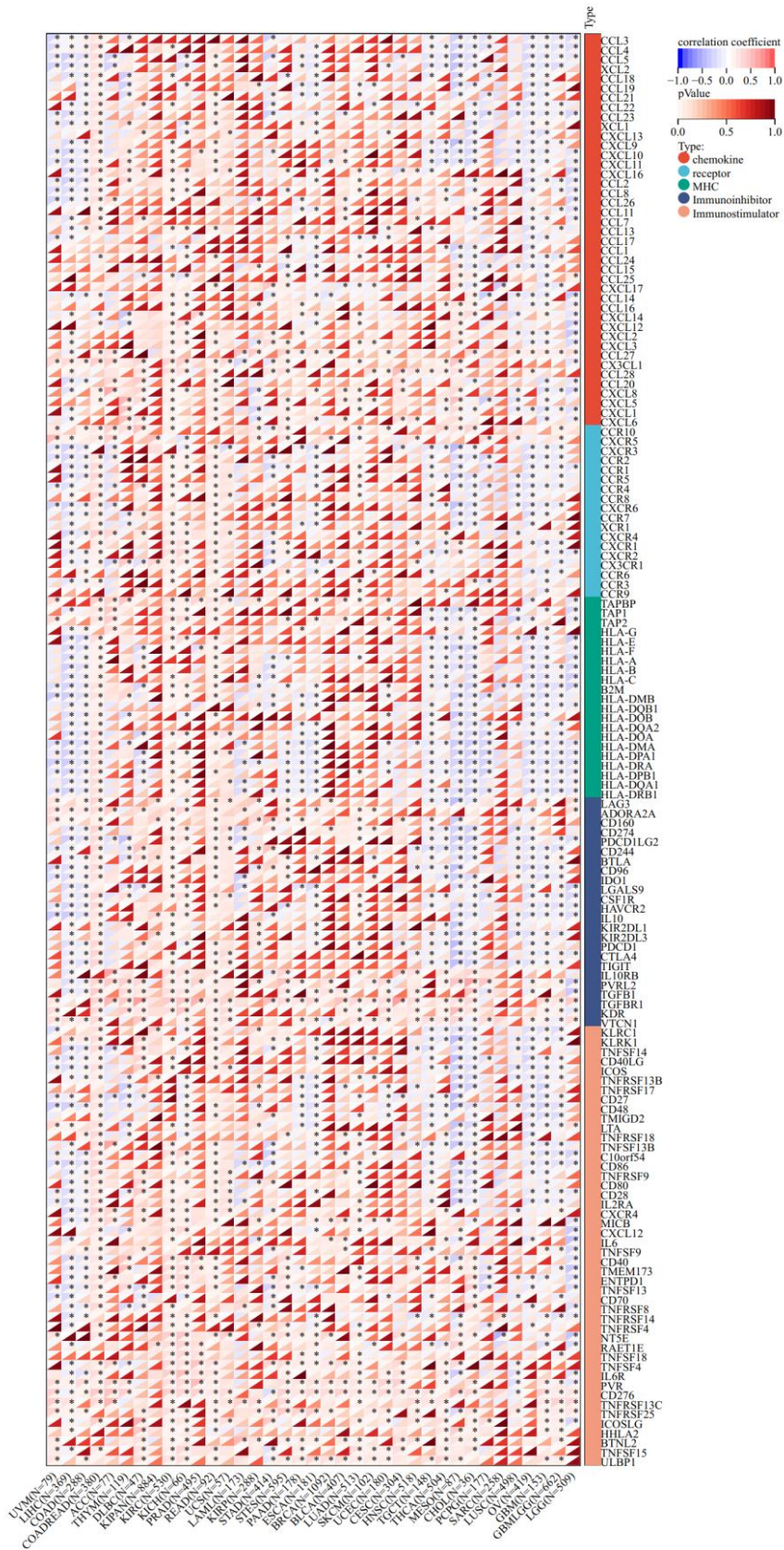

**FIGURE 11** Correlation analysis between RNMT expression and immunomodulators (chemokines, receptors, MHC, immunostimulators, and immunoinhibitors) in pan-cancer. (Spearman correlation, *p < 0.05.)

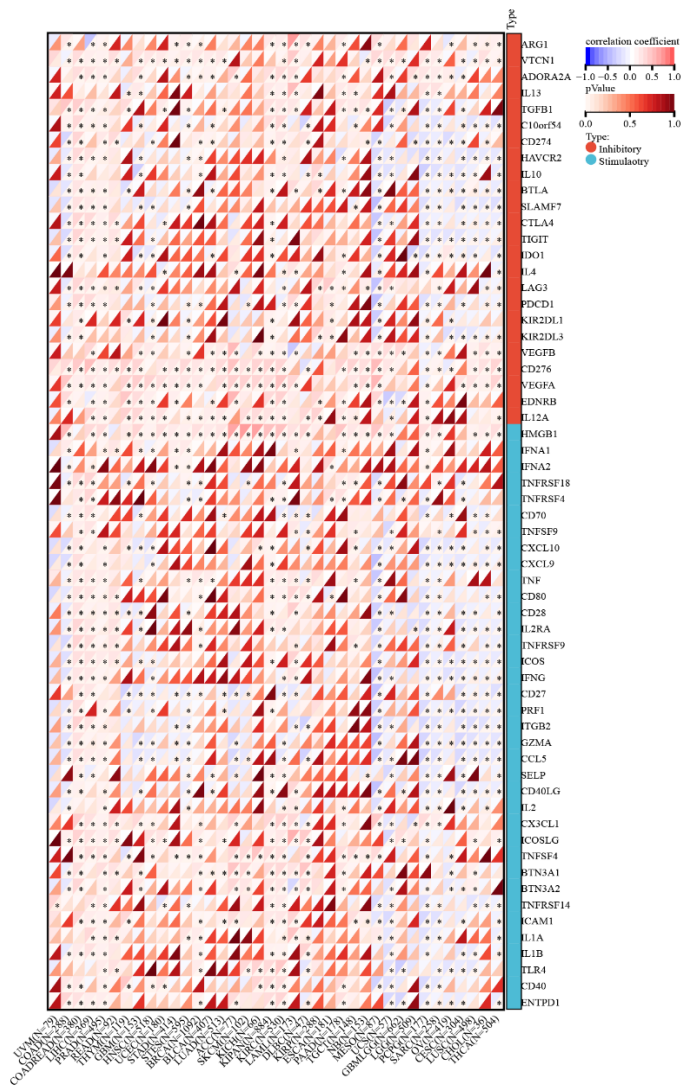

**FIGURE 12** Correlation analysis between RNMT expression and immune checkpoints (inhibitors and stimulators) in pan-cancer. (Spearman correlation, *$p < 0.05$.)

**Correlation Between RNMT Expression and TMB, MSI, DNA MMR Genes, and DNMT**

To better understand the role of RNMT, we analyzed the correlations between RNMT expression and TMB, MSI, DNA MMR genes, and DNMT in TCGA database. Our results showed that RNMT expression had significant positive associations with TMB in THYM, COAD, LAML, LUAD, SKCM, and STAD and negative relations in BRCA, PRAD, and THCA **(Figure 12B)**. For MSI, RNMT expression was found to be positively correlated in UCEC, CESC, COAD, LUSC, SARC, STAD, and TGCT and negatively correlated in DLBC, HNSC, PRAD, and THCA **(Figure 12A)**. Analyzing the relationships between RNMT and five MMR genes(MLH1, MSH2, MSH6, PMS2, and EPCAM), we found that RNMT expression was significantly correlated with mutations in MLH1, MSH2, MSH6, and PMS2. EPCAM showed a positive association with RNMT in BLCA, BRCA, HNSC, KICH, KIRC, KIRP, LAML, LIHC, LUAD, LUSC, OV, PAAD, PCPG, PRAD, STAD, TGCT, THCA, THYM, UCEC **(Figure 12C)**.

Then, we analyzed the correlation between RNMT expression and four methyltransferases (DNMT1, DNMT2, DNMT3A, and DNMT3B). Interestingly, we found that RNMT expression was strongly correlated with DNMT1, DNMT2, DNMT3A, and DNMT3B expression in human pan-cancer **(Figure 12D)**. In summary, these results indicate that RNMT expression was related to immunity and by modulating MMR and DNA methylation, it may regulate the progression and prognosis of tumors in different types of cancer.

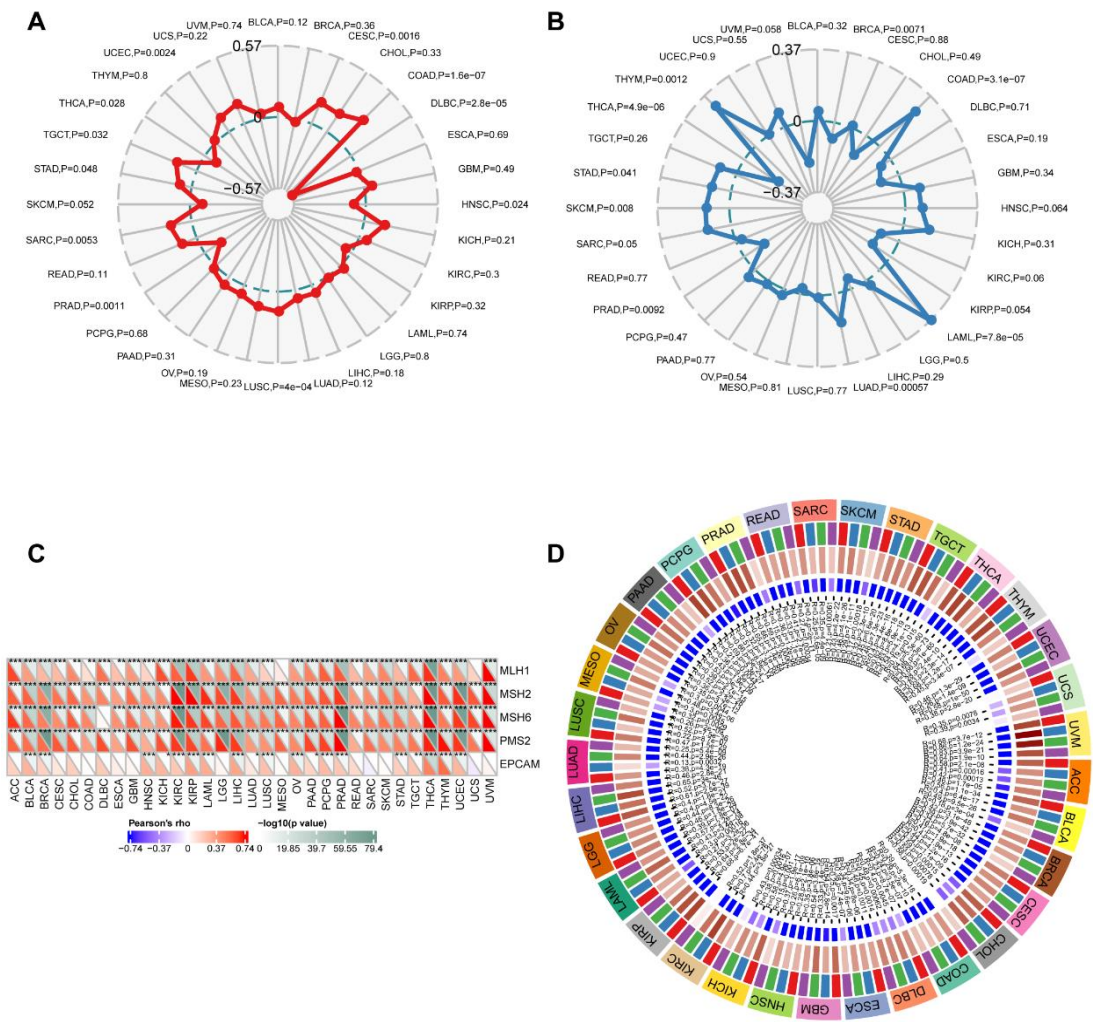

**FIGURE 12** Correlation analysis between RNMT expression and MSI **(A)**, TMB **(B)**, DNA MMR genes **(C)**, and DNMT **(D)** in pan-cancer. (DNMT1 was red, DNMT2 was blue, DNMT3 was green, and DNMT4 was purple. Spearman's correlation, $p < 0.05$ was considered significant, *$p < 0.05$, **$p < 0.01$, ***$p < 0.001$.)

**Functional analysis by GSEA**

To observe the effects of RNMT expression on different cancer types, GSEA was used to analyze the signaling pathways enriched in KEGG in high-expression and low-expression groups. According to the NES score permutation, the top 20 most enriched biological processes or signaling pathways have been previously characterized listed

**(Table 2)**. As shown in **Figure 13A**, RNMT was highly expressed in the mTOR signaling pathway, ubiquitin mediated proteolysis, and insulin signaling pathway. Notably, RNMT was highly expressed in the chronic myeloid leukemia, ERBB signaling pathway, prostate cancer, WNT signaling pathway, non-small cell lung cancer, pathway in cancer, endometrial cancer, and glioma. In contrast, RNMT was expressed in asthma, oxidative phosphorylation, parkinsons disease, and ribosome as the top 4 negatively enriched terms **(Figure 13B)**. These results indicate that RNMT widely regulates signaling pathways on immune and metabolic function.

**TABLE 2** The top 20 GSEA enrichment analysis were enriched in the high RNMT expression group in KEGG database.

| Term | ES | NES | NP | FDR | FWER |
|---|---|---|---|---|---|
| KEGG_MTOR_SIGNALING_PATHWAY | -0.6254 | -2.2118 | 0 | 0.0016 | 0.001 |
| KEGG_UBIQUITIN_MEDIATED_PROTEOLYSIS | -0.6075 | -2.1743 | 0 | 0.0035 | 0.003 |
| KEGG_INSULIN_SIGNALING_PATHWAY | -0.5253 | -2.1112 | 0 | 0.0076 | 0.01 |
| KEGG_CHRONIC_MYELOID_LEUKEMIA | -0.5875 | -2.0812 | 0 | 0.0074 | 0.014 |
| KEGG_ERBB_SIGNALING_PATHWAY | -0.5514 | -2.0803 | 0 | 0.0059 | 0.014 |
| KEGG_PROSTATE_CANCER | -0.5438 | -2.0455 | 0 | 0.0099 | 0.027 |
| KEGG_ENDOCYTOSIS | -0.5181 | -2.0408 | 0 | 0.0091 | 0.028 |
| KEGG_ADHERENS_JUNCTION | -0.5913 | -2.0142 | 0 | 0.0117 | 0.041 |
| KEGG_NEUROTROPHIN_SIGNALING_PATHWAY | -0.5346 | -2.0096 | 0 | 0.0116 | 0.046 |
| KEGG_TIGHT_JUNCTION | -0.5186 | -2.0028 | 0 | 0.0115 | 0.05 |
| KEGG_WNT_SIGNALING_PATHWAY | -0.4984 | -2.0016 | 0.002 | 0.0104 | 0.05 |
| KEGG_NON_SMALL_CELL_LUNG_CANCER | -0.562 | -1.9921 | 0 | 0.0111 | 0.057 |
| KEGG_LONG_TERM_DEPRESSION | -0.4895 | -1.9647 | 0 | 0.0151 | 0.07 |
| KEGG_INOSITOL_PHOSPHATE_METABOLISM | -0.552 | -1.9644 | 0 | 0.0143 | 0.07 |
| KEGG_GAP_JUNCTION | -0.513 | -1.9563 | 0 | 0.0146 | 0.077 |
| KEGG_LONG_TERM_POTENTIATION | -0.4973 | -1.9529 | 0 | 0.0146 | 0.084 |
| KEGG_PATHWAYS_IN_CANCER | -0.4601 | -1.9496 | 0 | 0.0143 | 0.089 |
| KEGG_DORSO_VENTRAL_AXIS_FORMATION | -0.6345 | -1.9355 | 0 | 0.0157 | 0.097 |
| KEGG_ENDOMETRIAL_CANCER | -0.565 | -1.9354 | 0 | 0.0149 | 0.097 |
| KEGG_GLIOMA | -0.5279 | -1.9314 | 0 | 0.0146 | 0.1 |

Enrichment score, ES; standardized enrichment score, NES; nominal p-value, NP; false discovery rate, FDR; Family-wise error rate, FWER.

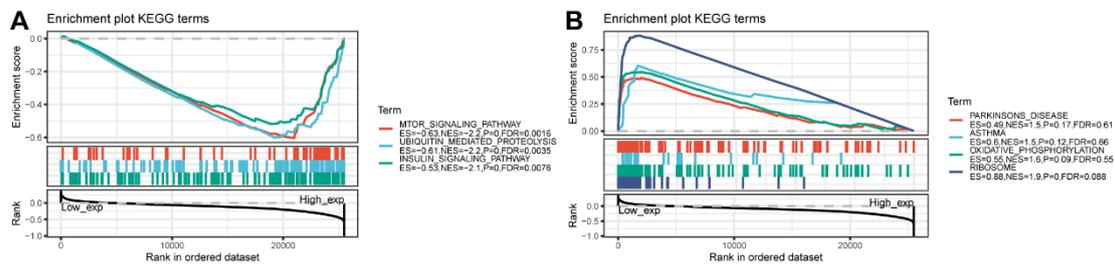

**FIGURE 13** KEGG analysis of the RNMT gene enrichment. **(A)** The top 3 terms were enriched in the high RNMT expression group. **(B)** The top 4 terms were enriched in the low RNMT expression group.

## DISCUSSION

Pan-cancer analysis can be used to investigate the similarities and differences among multiple cancer types.　More importantly, it can provide new clues and perspectives for cancer prevention and treatment (27). Recently, there are many studies focused on the pan-cancer analysis of the RNA alterations, revealing mutations, driver genes, and whole genome which are related to the occurrence and development of cancer (28-30).

Our study was intended to explore the dissimilarities of RNMT expression between normal tissues and tumors RNMT, hence we combined the samples of TCGA database and GTEx database to analyze. The result of a decreased expression in GBMLGG (21) and having no significant difference in BRCA were consistent with previous studies. However, Dunn S et al. identified that RNMT inhibition contributed to reducing PIK3CA mutant breast cancer cell proliferation (20). It suggested that RNMT may play a specific role in specific tumor types, and incorporated signaling pathway to explore might have an unexpected breakthrough.

Next, the correlation between RNMT expression and prognosis was researched. COX regression model and K-M curves gave a first indication that high RNMT expression patients had a poor prognosis in different cancer types, including LIHC, ACC, KICH, PRAD, STES, and LUSC. In contrast, up-regulated RNMT was correlated with good prognosis in GBMLGG, THYM, OV, and SKCM. These results proved that RNMT could represent a potential prognostic pan-cancer marker. Then, we demonstrated that the number of RNMT alternations was high in the majority of cancer types, which was consistent with previous research in colon cancer (31). Additionally, the report of Manning M et al (32) showed that more RNA methyltransferases (RNMTs) have high expressions in high-grade breast cancers and it indicated that RNMT probably played a considerable role in assessing the aggressiveness of cancers and more subtypes needed to be verified by in-depth experiments.

There is no separation between the genesis and development of tumor cells and the complex TME. The aberrant interactions between cellular components, such as T cell, B cell and NK cell affect the aggressiveness and degree of invasion of the tumors (33). According to this study, RNMT was positively related to various immune cell expressions in several types of tumors, therefore, it implied that RNMT was likely to be a promising and new regulator to improve TME.

Immunomodulators are molecules that can enhance or inhibit immune function or regulate both ways. For example, immunostimulators are predominantly used to improve the ability of anti-tumor and infection and correct immune deficiency. In addition, IL-2 is one of the most intensively studied cytokines in cancer immunotherapies for its effects in immunostimulation and immunosuppression, which has been approved by the FDA as a first-line treatment for patients with renal cell carcinoma and melanoma (34, 35). Immune checkpoints are a series of stimulation and inhibition pathways that prevent uncontrolled immune responses and modulate self-tolerance. They are associated with dysfunctional T cells and have the ability to regulate T cells (36). In general, immune checkpoints are involved in tumor immunosuppression and should be ideal modulation targets for tumor immunotherapy. In this study, we conducted a systematic analysis and found that RNMT was positively correlatied with immunostimulator CD 276 and TNFRSF13C. In particular, there is co-expression of RNMT with 24 immune inhibitors and 36 stimulators across cancers, specifically in UVM, COAD, COADREA, PRAD, KIPAN, and LIHC. The novel results of this study indicated that RNMT could serve as a key factor in cancer immunity by recruiting and regulating infiltrating immune cells to inhibit or promote the progression of cancers. Given that the expression of RNMT is significantly related to the prognosis of UVM, COAD, COADREA, PRAD, KIPAN, and LIHC, our research can provide a foundation for model animals and development of more immunotherapies in the future.

Furthermore, GSEA analysis of the pan-cancer study suggested that RNMT can participate in a variety of biosynthesis, metabolic and cancer pathways, especially in mTOR signaling pathway. Abnormal mTOR signaling pathway plays an important part during tumorigenesis and cancer development (37). In particular, a growing number of studies have revealed that mTOR signaling is associated with regulating immune infiltration in the TME (38, 39). For example, a recent study has shown that mTOR gene expression is associated with a variety of immune cells and immunoinhibitors in clear cell renal cell carcinoma (ccRCC) (40). Additionally, mTOR signaling pathway is associated with the development and functional properties of Th17, memory CD8 + T-cells, and NK cells (41). The aforementioned studies indicated that the interaction between RNMT and GSEA signaling pathway in TME may be a feasible study direction.

Overall, our study demonstrated that RNMT was differently expressed in normal and cancer samples, and associated with various prognoses in different cancer types. Furthermore, RNMT alternations were related to tumorigenesis and the development of tumors. On this basis, we revealed that RNMT was associated with immune and molecular subtypes, tumor immune infiltration and microenvironment, immunoregulators, MMR, DNA methylation, MSI, and TMB in numerous cancers. RNMT was involved in various pathways that influenced tumor immune and the development of tumors. The mentioned study suggested that RNMT may be a potential therapeutic target and immunological biomarker in various cancers. However, our study was based on public databases, and there's no experimental validation, but we have to acknowledge that RNMT has a strong association with different cancer types in cancer development and tumor immunity. RNMT holds promise as a potential biomarker for various cancer types.